\documentclass[twoside,journey]{IEEEtran}
\usepackage{makecell}
\usepackage{hyperref}
\usepackage{array}
\usepackage{graphicx,amssymb,amsmath}
\usepackage{multicol}
\usepackage[noadjust]{cite}
\usepackage{setspace}
\usepackage{subfigure}
\usepackage{float}
\usepackage {url}
\usepackage{stfloats}
\usepackage{amsthm,pifont}
\usepackage{flushend}
\usepackage{cases,subeqnarray}
\usepackage{bm,multirow,bigstrut}
\usepackage{amsmath, amsthm, amssymb}
\usepackage{textcomp}
\usepackage{latexsym,bm}
\usepackage{booktabs}
\usepackage{xcolor}
\usepackage{mathtools}
\usepackage{dsfont}
\usepackage{extarrows}
\usepackage{epsfig}
\usepackage{epsfig}
\usepackage{epstopdf}
\usepackage{colortbl}
\usepackage[noend]{algpseudocode}
\usepackage{algorithmicx,algorithm}

\theoremstyle{plain}

\theoremstyle{plain}

\newtheorem{prop}{Proposition}
\usepackage{amsmath}

\definecolor{Gray}{gray}{0.85}

\IEEEoverridecommandlockouts
\begin{document}

\title{Mixture of Experts for Network Optimization:\\
A Large Language Model-enabled Approach}
\author{Hongyang~Du, Guangyuan~Liu, Yijing~Lin, Dusit~Niyato,~\IEEEmembership{Fellow,~IEEE}, Jiawen~Kang, Zehui~Xiong, and Dong~In~Kim,~\emph{Fellow,~IEEE}
\thanks{H.~Du, G.~Liu, and D.~Niyato are with the School of Computer Science and Engineering, Nanyang Technological University, Singapore 639798, Singapore (e-mail: hongyang001@e.ntu.edu.sg; liug0022@e.ntu.edu.sg; dniyato@ntu.edu.sg). Y.~Lin is with the State Key Laboratory of Networking and Switching Technology, Beijing University of Posts and Telecommunications, 100876, Beijing, China (e-mail: yjlin@bupt.edu.cn).
J.~Kang is with the School of Automation, Guangdong University of Technology, and Key Laboratory of Intelligent Information Processing and System Integration of IoT, Ministry of Education, Guangzhou 510006, China, and also with Guangdong-HongKong-Macao Joint Laboratory for Smart Discrete Manufacturing, Guangzhou 510006, China (e-mail: kavinkang@gdut.edu.cn). Z.~Xiong is with the Pillar of Information Systems Technology and Design, Singapore University of Technology and Design, Singapore 487372, Singapore (e-mail: zehui\_xiong@sutd.edu.sg). D.~I.~Kim is with the Department of Electrical and Computer Engineering, Sungkyunkwan University, Suwon 16419, South Korea (e-mail: dikim@skku.ac.kr).}
}
\maketitle
\vspace{-1cm}

\begin{abstract}
Optimizing various wireless user tasks poses a significant challenge for networking systems because of the expanding range of user requirements. Despite advancements in Deep Reinforcement Learning (DRL), the need for customized optimization tasks for individual users complicates developing and applying numerous DRL models, leading to substantial computation resource and energy consumption and can lead to inconsistent outcomes. To address this issue, we propose a novel approach utilizing a Mixture of Experts (MoE) framework, augmented with Large Language Models (LLMs), to analyze user objectives and constraints effectively, select specialized DRL experts, and weigh each decision from the participating experts. Specifically, we develop a gate network to oversee the expert models, allowing a collective of experts to tackle a wide array of new tasks. Furthermore, we innovatively substitute the traditional gate network with an LLM, leveraging its advanced reasoning capabilities to manage expert model selection for joint decisions. Our proposed method reduces the need to train new DRL models for each unique optimization problem, decreasing energy consumption and AI model implementation costs. The LLM-enabled MoE approach is validated through a general maze navigation task and a specific network service provider utility maximization task, demonstrating its effectiveness and practical applicability in optimizing complex networking systems.
\end{abstract}
\IEEEpeerreviewmaketitle

\section{Introduction}
As we step into the era of Sixth-Generation (6G) networks, the dynamics of wireless communication and network systems are undergoing significant transformation, propelled by increasing complexity and a widening array of user needs~\cite{dang2020should}. These advanced networks are anticipated to offer unparalleled speed and connectivity while being fundamentally user-focused, flexible, and smart~\cite{ammar2021user}. Consequently, the need for sophisticated optimization within network systems has intensified, becoming essential to unlock the extensive capabilities of 6G. Among various technological innovations, Deep Reinforcement Learning (DRL) is a critical enabler~\cite{luong2019applications}. DRL's inherent flexibility and ongoing learning potential make it exceptionally capable of meeting dynamic demands of evolving networks, effectively responding to the complex patterns of user interactions and requirements~\cite{luong2019applications}.

\begin{figure}[t]
\centering
\includegraphics[width = 0.38\textwidth]{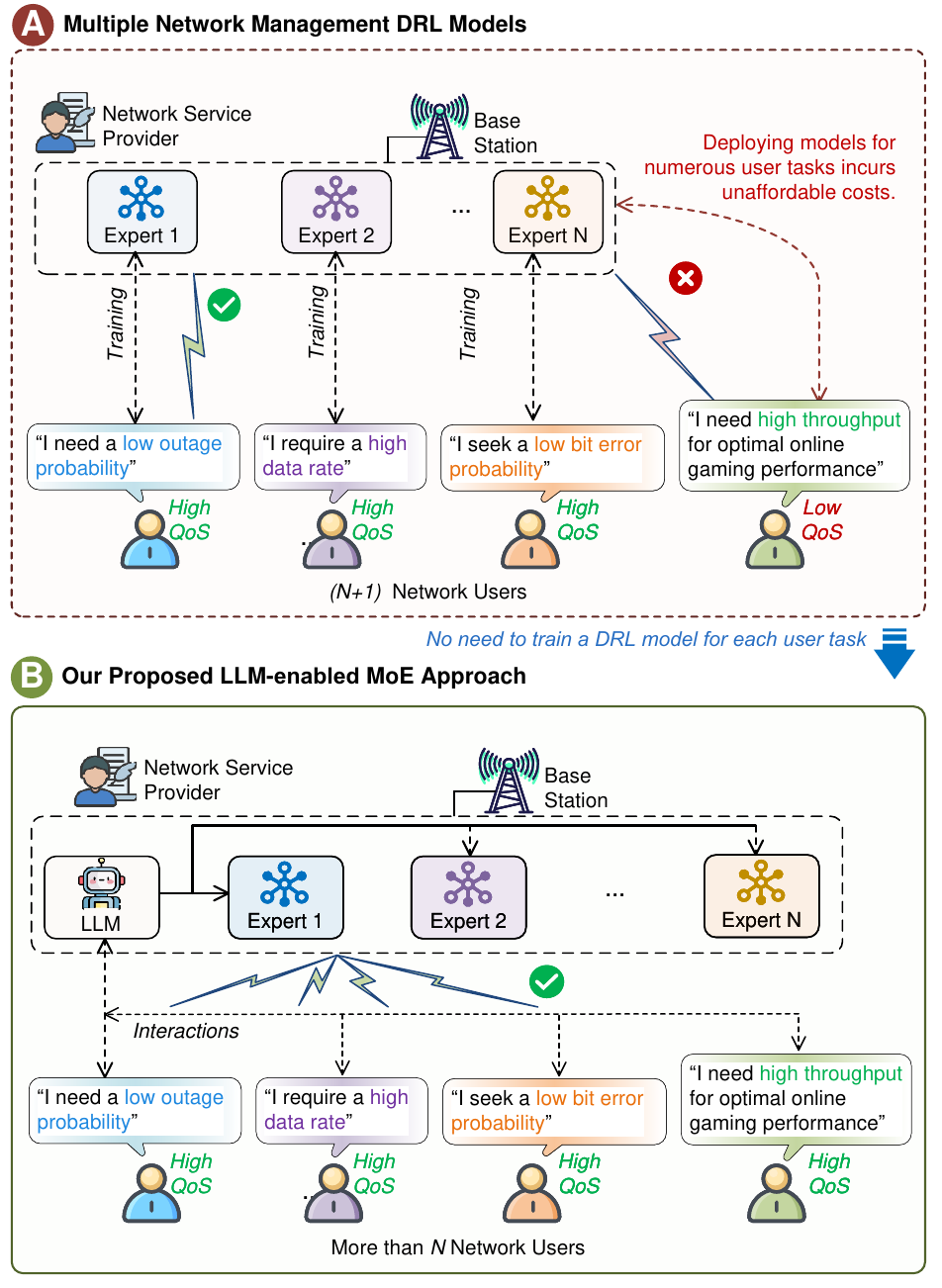}
\caption{Network optimization strategies. {\textbf{Part A}} demonstrates the drawbacks of training distinct AI models for different user requirements, emphasizing the costs of excessive AI model deployment. {\textbf{Part B}} presents our LLM-enabled MoE approach, using a limited set of DRL models to efficiently address a variety of user tasks.}
\label{motivation}
\end{figure}
However, the rapid expansion of DRL models, each tailored for specific tasks, significantly strains network servers due to the high demands for training and deployment resources. As shown in Fig.~\ref{motivation}, the pre-trained DRL models are inadequate when a user presents a new requirement. Consequently, this situation presents a significant challenge in pursuing user-centric networks. A critical question arises:
\begin{itemize}
\item \textit{How can we achieve effective network optimization without using numerous DRL models individually trained for each specific task?}
\end{itemize}
Recent advancement of the Mixture of Experts (MoE) framework offers an effective solution~\cite{du2023user}. By employing a range of AI models as specialized experts, MoE supports collaborative decision-making, significantly reducing the need for individual task-specific model training. Within this framework, actor networks, trained by varied DRL policies, can act as expert models to tackle new and complex user tasks by jointly making decisions. 
A gate network is conventionally trained to manage and schedule these expert models, ensuring optimal task handling. 
However, training the gate network presents its own set of challenges, including uncertainties in performance and potential limitations imposed by the number of experts and task complexity. 

\begin{figure}[t]
\centering
\includegraphics[width = 0.4\textwidth]{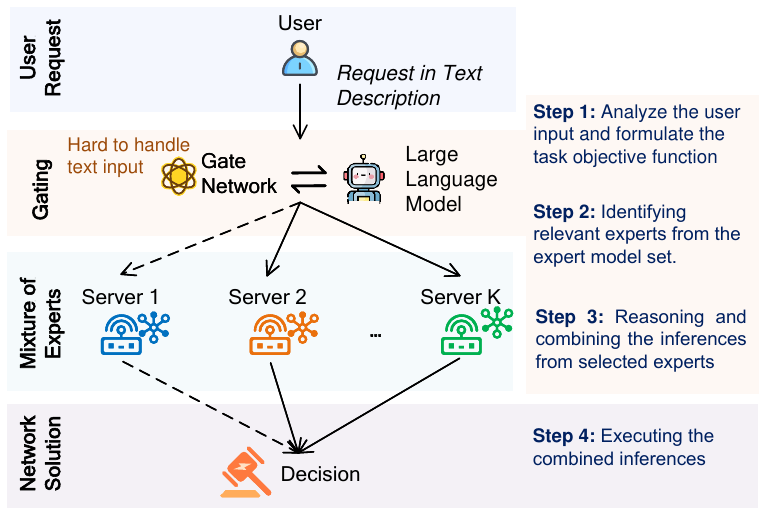}
\caption{Workflow of the proposed LLM-enabled MoE framework: Upon receiving a text-based description of their requirements, the LLM processes and reasons about the user's needs, identifying the experts necessary for the task at hand and determining their decision-making weights.}
\label{framework}
\end{figure}

Fortunately, the advancement in Large Language Models (LLMs) presents a promising solution. 
As shown in Fig.~\ref{framework}, LLMs, with their extensive knowledge bases and strong reasoning capabilities, are adept at understanding user requirements through text-based interactions~\cite{zhao2023survey,du2023user}. 
This capability to accurately interpret and react to user inputs positions LLMs as suitable alternatives for the gate network in the Mixture of Experts (MoE) architecture.
As a result, LLMs can effectively orchestrate the selection and integration of specialized expert models, thereby improving the system's decision-making efficacy and responsiveness to user requirements~\cite{du2023user}.
Thus, in this paper, we propose an innovative approach for optimizing user-centric network systems by leveraging the advanced capabilities of LLMs alongside the MoE framework. 
Within the LLM-enabled MoE framework, each expert model is a distinct DRL model deployed on separate edge servers optimized for specific network tasks. 
The LLM facilitates the alignment of DRL model outputs with user requirements, improving the collective decision-making mechanism. We summarize our main contributions as follows:
\begin{itemize}
\item By adopting the MoE framework, we facilitate the cooperative operation of various DRL models, improving the network's efficiency and capacity to adapt to evolving user requirements.
\item We integrate LLMs into the MoE architecture, enabling a synergistic approach where multiple DRL models work together under the guidance of an LLM to address new network optimization problems.
\item We evaluate our LLM-enabled MoE approach through empirical testing on a standard DRL task, i.e., maze navigation, and a network optimization task to maximize utility for Network Service Providers (NSPs). These tests have demonstrated the effectiveness and versatility of our approach in practical scenarios.
\end{itemize}
A list of mathematical symbols frequently used in this paper is available in Table I.
\begin{table}
\caption{Key Mathematical Notations.}
\vspace{-0.1cm}
\label{Networkparameters}
\centering
\renewcommand{\arraystretch}{1}
{\small\begin{tabular}{m{1.5cm}<{\centering}|m{6cm}<{\centering}}
\toprule
\hline
\textbf{Notation} & \textbf{Description} \\
\hline
${\bm{s}}_k$ & The requirement of the $k_{\rm th}$ user \\
\hline
${\bf{o}}_k$ &  The task objective \\
\hline
${\bf{m}}_{\rm all}$ &  The set of available expert models for the LLM \\
\hline
${\bf{m}}_{k}$ & The selected expert models \\
\hline
${\bf{g}}_k$ & The additional information about the network optimization problem \\
\hline
${\bf{d}}_{k}$ & The final decision\\
\hline
$M$ &  Number of antennas in BS \\
\hline
$K$ & Number of users  \\
\hline
$N$ & Number of DRL models \\
\hline
$D_{k}$ & Transmission distance of the $k_{\rm th}$ BS-user pair  \\
\hline
$\alpha_{k}$ & Path loss exponent of the $k_{\rm th}$ BS-user pair  \\
\hline
$P_{k}$ & Downlink transmit power  \\
\hline
\bottomrule
\end{tabular}}
\end{table}

\section{LLM-enabled Mixture-of-Experts Approach}
This section introduces the LLM-enabled MoE approach by considering a general DRL environment, i.e., the grid-world maze. Notably, this example can seamlessly extend to a wide range of network optimization problems, such as optimal service network selection, load balancing, and power allocation.

\subsection{Environment Settings}
\begin{figure*}[t]
\centering
\includegraphics[width = 0.8\textwidth]{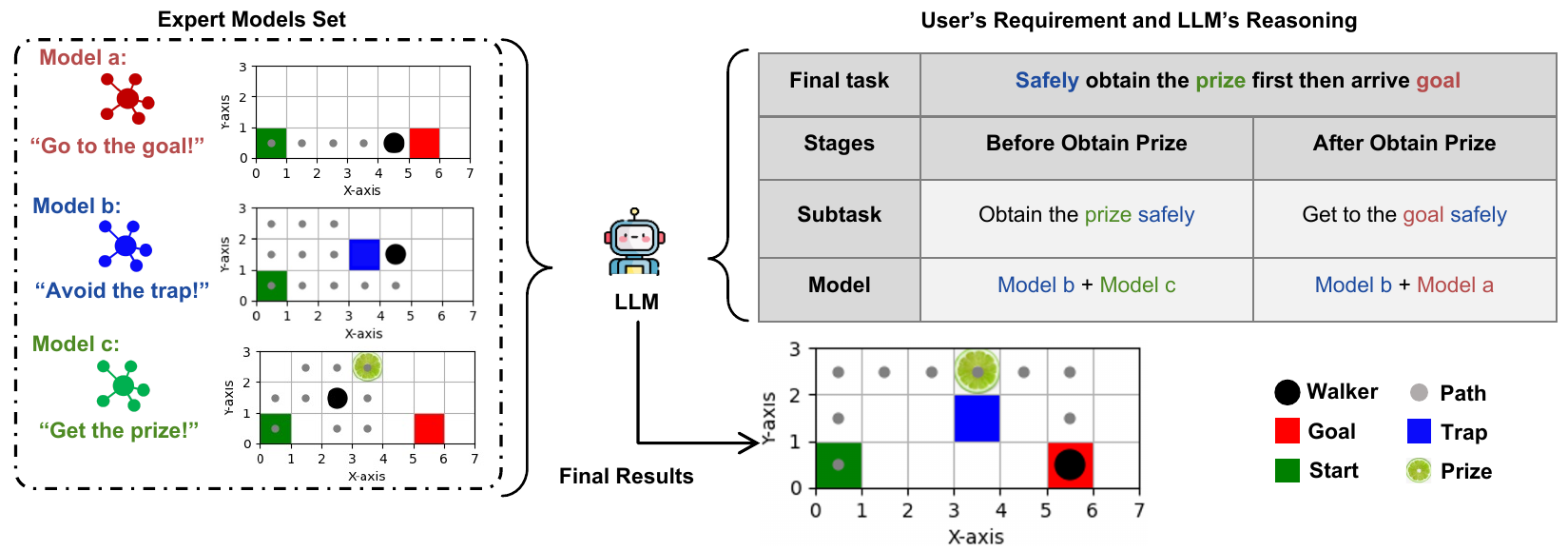}
\vspace{-0.2cm}
\caption{LLM-enabled MoE framework demonstration in a maze navigation task. An ensemble of DRL models trained on diverse tasks serves as a set of expert models accessible to the LLM. The LLM analyzes and infers user tasks, leveraging combinations of expert models to address the final objectives.}
\label{DRLT}
\end{figure*}
The grid-world maze \cite{kaelbling1996reinforcement} is set up as a $3$ $\times$ $7$ grid where a walker navigates to achieve specific requirements. In the maze, there are several special positions:
\begin{itemize}
\item \textit{Goal:} The destination that the walker aims to reach.
\item \textit{Prize:} A designated grid position that awards a positive reward upon the walker's arrival.
\item \textit{Trap:} A grid position that incurs a penalty when encountered by the walker.
\end{itemize}
The tasks in this grid-world maze environment are diverse, aiming not only to reach specified goals but also to collect prizes and avoid traps. Users can define these tasks through textual commands, such as {\textit{``Find the shortest path to the goal,''}} or {\textit{``Navigate the walker to explore the environment and obtain the prize.''}} The diversity of tasks in this environment mirrors similar scenarios in network optimization, where user requirements can vary widely, e.g., {\textit{``Optimize bandwidth to improve streaming quality,''}} or {\textit{``Balance network load to prevent congestion during peak hours.''}}

\subsection{LLM-enabled MoE Approach}
In the LLM-enabled MoE framework, the gate network is trained via specific algorithms, like reinforcement or supervised learning, to ascertain the applicability of diverse expert models to user-defined requirements. For instance, to train a gate network in the maze navigation task using the DRL, we define the {\textit{state}} as the walker's precise location on the grid. The {\textit{action}}, represented as a vector, determines each expert model's contribution weight, where a zero value negates the respective expert's influence.
The {\textit{reward}} mechanism is defined according to user requirements to ensure the gate network's behavior aligns with its final goals. 
However, the mathematical expression of {\textit{reward}} can vary significantly based on user needs, presenting a significant challenge. 
Herein lies the opportunity for integrating LLMs. With the analytical prowess of LLMs, it is feasible to infer the necessary expert models and deduce their combinatory decision-making process directly from user requirements without additional training.
As shown in Fig.~\ref{DRLT}, the LLM evaluates the current state, objectives, and available set of expert models, selecting suitable experts to achieve the final task. The LLM-enabled MoE inference process involves:
\begin{itemize}
\item {\textit{Step 1: Objective Formulation:}} Analyzing user input and system state to formulate the task objective function.
\begin{equation}
{\text{LLM}}\{{\bm{s}}_k\} \to {\bf{o}}_k,
\end{equation}
where ${\text{LLM}}\{\cdot\}$ denotes the inference process of the LLM, ${\bm{s}}_k$ is the requirement of the $k_{\rm th}$ user such as {\textit{``I want to arrive the goal in the safest way''}}, and ${\bf{o}}_k$ is the task objective such as {\textit{\{``Go to the goal''}}, {\textit{``Avoid the trap''\}}}.
\item \textit{Step 2: Expert Selection:} Identifying relevant experts from the set of expert models based on the formulated objective. 
\begin{equation}
{\text{LLM}}\{{\bf{o}}_k, {\bf{m}}_{\rm all}\} \to {\bf{m}}_{k},
\end{equation}
where ${\bf{m}}_{\rm all}$ is the set of expert models and ${\bf{m}}_{k}$ is the selected models. For example, for ${\bf{o}}_k$ in Step 1, the expert models $a$ and $b$ are selected to perform the task.
\item \textit{Step 3: Inference Combination:} Reasoning and combining the inferences from selected experts.
\begin{equation}
{\text{LLM}}\{{\bf{m}}_{k}, {\bf{o}}_k, {\bf{g}}_k\} \to {\bf{d}}_{k},
\end{equation}
where ${\bf{g}}_k$ is the additional information about the network optimization problem, such as the wireless network conditions, and ${\bf{d}}_{k}$ is the final decision, i.e., the path of the walker.
\item \textit{Step 4: Decision Execution:} Executing the combined inferences, i.e., ${\bf{d}}_{k}$.
\end{itemize}

\section{Applications in Intelligent Networks}
This section examines a utility maximization problem for (NSPs), where users exhibit varying Quality of Service (QoS) demands. We show the functionality of our proposed LLM-enabled MoE approach in addressing the NSPs' utility maximization problem under users' diverse requirements.
\begin{figure}[t]
\centering
\includegraphics[width = 0.4\textwidth]{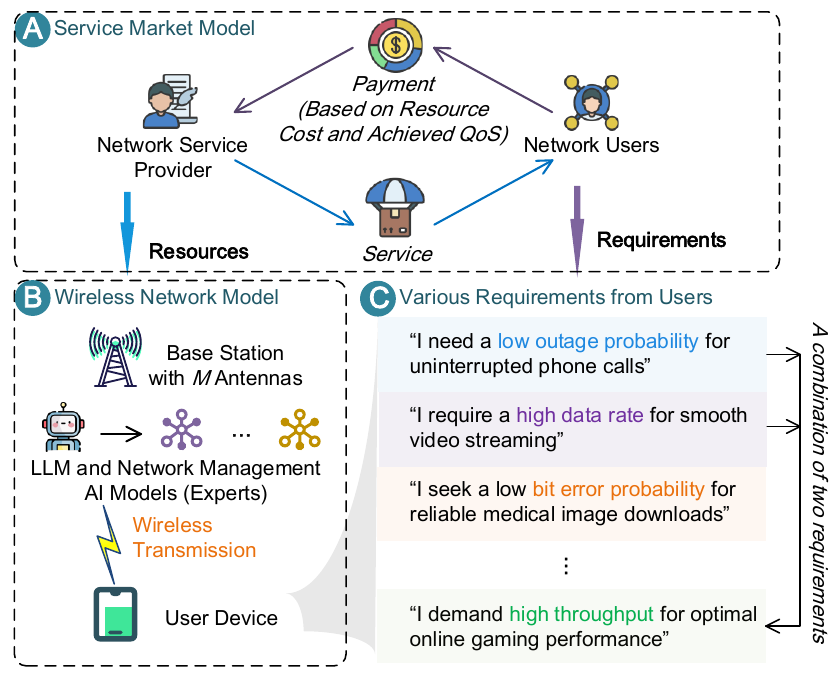}
\vspace{-0.2cm}
\caption{System model. {\textbf{Part A}} shows the {\textit{Service Market Model}}, illustrating the interaction between the {\textit{NSPs}} and {\textit{Users}}. {\textbf{Part B}} represents the {\textit{Wireless Network Model}}, wherein we consider a BS with $M$ antennas providing services to a user device. {\textbf{Part C}} shows various user requirements under different scenarios, which affect the payment structure in {\textbf{Part A}} and the optimal power allocation strategies in {\textbf{Part B}}.}
\label{systemmodel}
\end{figure}

\subsection{System Model}
As shown in \textbf{Part A} of Fig.~\ref{systemmodel}, we consider that the NSP provides a range of services, including voice calls, video streaming, and image downloading, to users via a wireless network. Within this framework, users, i.e., direct end-users and subscription-based intermediaries that consolidate demand, subscribe the NSP based on their specific QoS requirements, as depicted in {\textbf{Part C}} of Fig.~\ref{systemmodel}.
We consider the \textit{service market model} between NSPs and $K$ users~\cite{nguyen2018contract}. Specifically, the NSPs offer network resources, i.e., transmit power $P_k$ $\left( k = 1, \ldots, K\right)$, in exchange for payment. The users have different requirements that define their expected QoS, ranging from {\textit{low Outage Probability (OP)}} for uninterrupted voice calls to {\textit{high throughput}} for online gaming performance. Let ${\bm{g}}_k$ and ${{\cal Q}_k}$ denote the $k_{\rm th}$ user's wireless conditions and QoS, respectively. We define the optimization problem for each $k_{\rm th}$ BS-user pair as follows:
\begin{equation}\label{opteq}
\begin{array}{*{20}{l}}
{\mathop {\max }\limits_{ {{P_k}} } }&{\beta_1 {{\mathcal{F}}\left({{\cal Q}_k}\left( {{P_k},{{\bm s}_k}, {\bm g}_k} \right)\right)}  - \beta_2 {P_k}}\\
{\quad\:{\rm{s.t.}},}&{{P_k} \le {P_{{\rm{th}}}},}\\
{}&{P_k \in \{P_1^{(a)},\ldots, P_L^{(a)}\}},\\
{}&{{{\cal Q}_k}\left( {{P_k},{{\bm s}_k}, {\bm g}_k} \right) \ge {Q_{k,\min }}}
\end{array}
\end{equation}
where $\beta_1$ represents the unit received payment from the users, ${{\cal Q}_{k,\min }}$ is the lowest QoS that the $k_{\rm th}$ user can accept, ${\mathcal{F}}\left( \cdot \right)$ is a utility function that maps diverse QoS metrics to a standardized assessment framework of user satisfaction and corresponding charges, $\beta_2$ denotes the cost coefficient, reflecting the resource expenditure of the NSP for transmit power, $\{P_1^{(a)},\ldots, P_L^{(a)}\}$ is a set of available power value settings, and $P_{\rm th}$ is the transmit power threshold.

We consider that the ${{\cal Q}_{k,\max }}$ to be the upper bound on the performance metric that the network can provide to the $k_{\rm th}$ user. Thus, ${\mathcal{F}}\left( \cdot \right)$ can be modeled as~\cite{du2023attention}
\begin{equation}
{\mathcal{F}}\left( {{{\cal Q}_k}\left( {{P_k},{{\bm s}_k}, {\bm g}_k} \right)} \right) = \frac{{{\cal Q}_k\left( {{P_k},{{\bm s}_k}, {\bm g}_k} \right) - {{\cal Q}_{k,\min }}}}{{{{\cal Q}_{k,\max }} - {{\cal Q}_{k,\min }}}}.
\end{equation}
To effectively tackle the optimization challenge in~\eqref{opteq}, defining the QoS formulation based on users' requirements is crucial. For instance, a textual request such as {\textit{``I am making a call and need to ensure continuity''}} implies that QoS primarily focuses on OP, whereas {\textit{``I am downloading medical images, accuracy is critical''}} suggests that Bit Error Probability (BEP) is the main QoS concern. This understanding allows for tailored optimization approaches that align with specific user needs, enhancing the overall service efficacy.

\subsection{Set of Expert Models Training}
Various DRL models can be trained according to specific QoS requirements, which form a set of expert models. Here, we consider two representative scenarios:
\begin{itemize}
\item \textbf{Case 1. OP:} Users require a \textit{low OP} to ensure uninterrupted voice calls, which is paramount for both professional and personal communications. Here, we have ${{\cal Q}_k}\left( {{P_k},{{\bm s}_k}, {\bm g}_k} \right) = 1 - {\rm{OP}}$.
\item \textbf{Case 2. Data Rate (DR):} A \textit{high DR} is essential for users using video streaming, enabling a buffer-free experience with high-definition content. In this case, we have ${{\cal Q}_k}\left( {{P_k},{{\bm s}_k}, {\bm g}_k} \right) = {\rm{DR}}$.
\end{itemize}

As depicted in \textbf{Part B} of Fig.~\ref{systemmodel}, we consider a multiple antenna Base Station (BS) employed for service delivery within the wireless network. 
For the $k_{\rm th}$ BS-user pair, the baseband received signal at each symbol period can be expressed as ${\bf{r}} = \sqrt{{D}^{-\alpha}}{\bf{H}}{\bf{w}}x + {\bf{n}}$, where \(\mathbf{H}\) represents the channel matrix with elements denoting the channel gains, \(\mathbf{w}\) is the weight vector applied at the transmitter, \(x\) denotes the transmitted symbols, and \(\mathbf{n}\) is the noise vector at the receiver. 
The weight vector, \(\mathbf{w}\) for the Maximum Ratio Transmission (MRT) is designed to align with the conjugate of the channel matrix \(\mathbf{H}\), thus \(\mathbf{w} = P \mathbf{H}^H / {\|\mathbf{H}\|}\), where \(\|\mathbf{H}\|\) denotes the Frobenius norm of \(\mathbf{H}\).
The SNR under the MRT scheme is formulated as follows:
\begin{equation}
{\rm{SNR}} = \frac{{{D^{ - \alpha }}{{\left\| {{\bf{Hw}}s} \right\|}^2}}}{{{\sigma ^2}}} = \frac{{P{D^{ - \alpha }}\sum\limits_{j = 1}^M {h_{k,j}^2} }}{{{\sigma ^2}}},
\end{equation}
where \(P\) represents the total transmit power, and \(\sigma^2\) denotes the noise power. 
Considering each \(h_j\) in \(\mathbf{H}\) follows a Rayleigh distribution, the squared magnitude \(|h_j|^2\) adheres to an exponential distribution. The Probability Density Function (PDF) of the effective channel gain, $Y = \sum\limits_{j = 1}^M {h_{k,j}^2}$, can be then given by $f_{{\rm{Y}}}(y) = \frac{y^{M-1}}{\Gamma\left(M\right)\theta^M} e^{-\frac{y}{\theta}}$, where \( \theta \) is the mean power of the sum of squared channel gains and \( \Gamma(\cdot) \) is the Gamma function. The PDF of \( \text{SNR} \) is obtained by scaling and transforming the PDF of \( Y \), leading to:
\begin{equation}\label{PDFeq}
f_{\text{SNR}}(z) = \frac{(\frac{\sigma^2}{P {D^{ - \alpha }} })^M z^{M-1}}{\Gamma(M)\theta^M} e^{-\frac{\sigma^2 z}{\theta P{D^{ - \alpha }}}},
\end{equation}
where \( z \) is the SNR variable.
We then explore the mathematical formulation of the reward function of the DRL models by deriving the closed-form of network performance metrics:
\begin{prop}\label{prop1}
The OP can be derived as follows:
\begin{equation}\label{opequation}
{\rm{OP}} = \frac{{\Gamma \left( {M,\frac{{{\sigma ^2}}}{{\theta P{D^{ - \alpha }}}}{\gamma _{{\rm{th}}}}} \right)}}{{\Gamma \left( M \right)}},
\end{equation}
where $\gamma_{\rm th}$ is the threshold for communications outage, and $\Gamma\left(\cdot,\cdot\right)$ is the upper incomplete Gamma function \cite[eq. (8.350.2)]{gradshteyn2007}.
\end{prop}

\begin{IEEEproof}
Please refer to Appendix~\ref{app1}.
\end{IEEEproof}

\begin{prop}\label{prop2}
The achievable DR for given bandwidth $B$ and SNR can be expressed as:
\begin{equation}\label{drequation}
{\rm{DR}} = \frac{B}{{\ln \left( 2 \right)\Gamma \left( M \right)}}G_{2,3}^{3,1}\left( {\left. {\frac{{{D^\alpha }{\sigma ^2}}}{{\theta P}}} \right|\begin{array}{*{20}{c}}
{0,1}\\
{M,0,0}
\end{array}} \right).
\end{equation}
\end{prop}
\begin{IEEEproof}
Please refer to Appendix~\ref{app2}.
\end{IEEEproof}
Leveraging {\textbf{Propositions 1 and 2}}, NSP utility maximization models, taking into account user QoS metrics such as OP and DR, can be effectively trained. Similarly, DRL models can be developed for other QoS indicators, such as BEP. These models form the set of expert models, which the LLM then utilizes for optimization tasks.

\subsection{LLM-enabled MoE for New User Requirement}
We consider an optimization task that arises from a new user requirement for which no dedicated DRL model has been previously trained:
\begin{itemize}
    \item \textbf{Case 3. New User Requirement:} Users engaged in online games require \textit{smooth experience} with the requirement: ${\bm s}_k$ $=$ \textit{``I need seamless and uninterrupted gaming sessions''}.
\end{itemize}

Given this new requirement, we explore the application of the LLM-enabled MoE framework as follows:
\begin{itemize}
    \item \textit{Step 1: Objective Formulation:} The LLM begins by interpreting the user's input ${\bm s}_k$, to establish task objectives, transforming user needs into actionable goals.
    The output of LLM\footnote{We show the key part of LLM's output due to the page limitation.} is {\textit{`` `Seamless' implies a smooth and uninterrupted flow of data transmission, and `uninterrupted' suggests that the flow of bits is continuous and consistent''. Considering these factors, it is reasonable to consider throughput as a key aspect. Throughput directly impacts the speed and efficiency of data transfer, affecting gameplay's responsiveness and overall smoothness.''}} Note that the throughput, i.e., ${\rm TP}$, is defined as
    \begin{equation}
        {\rm TP} = \left( 1 - {\rm OP} \right) {\rm DR}.
    \end{equation}
    \item \textit{Step 2: Expert Selection:} Without a retrained DRL model for throughput, the LLM assesses the available set of expert models to identify those relevant to achieving high throughput. The output of LLM is {\textit{``To optimize throughput for gaming sessions, the focus is on minimizing OP and maximizing DR.''}}
    \item \textit{Step 3: Inference Combination:} The LLM then synthesizes the selected experts' inferences, reasoning out how best to combine their outputs to enhance throughput, factoring in the minimization of OP and maximization of DR. More analysis is given in~Section \ref{fheliah}.
    \item \textit{Step 4: Decision Execution:} Finally, the network executes the derived transmit power setting.
\end{itemize}

\section{Numerical Analysis}
\subsection{Maze Navigation Task}
\begin{figure*}[t]
\centering
\includegraphics[width = 0.8\textwidth]{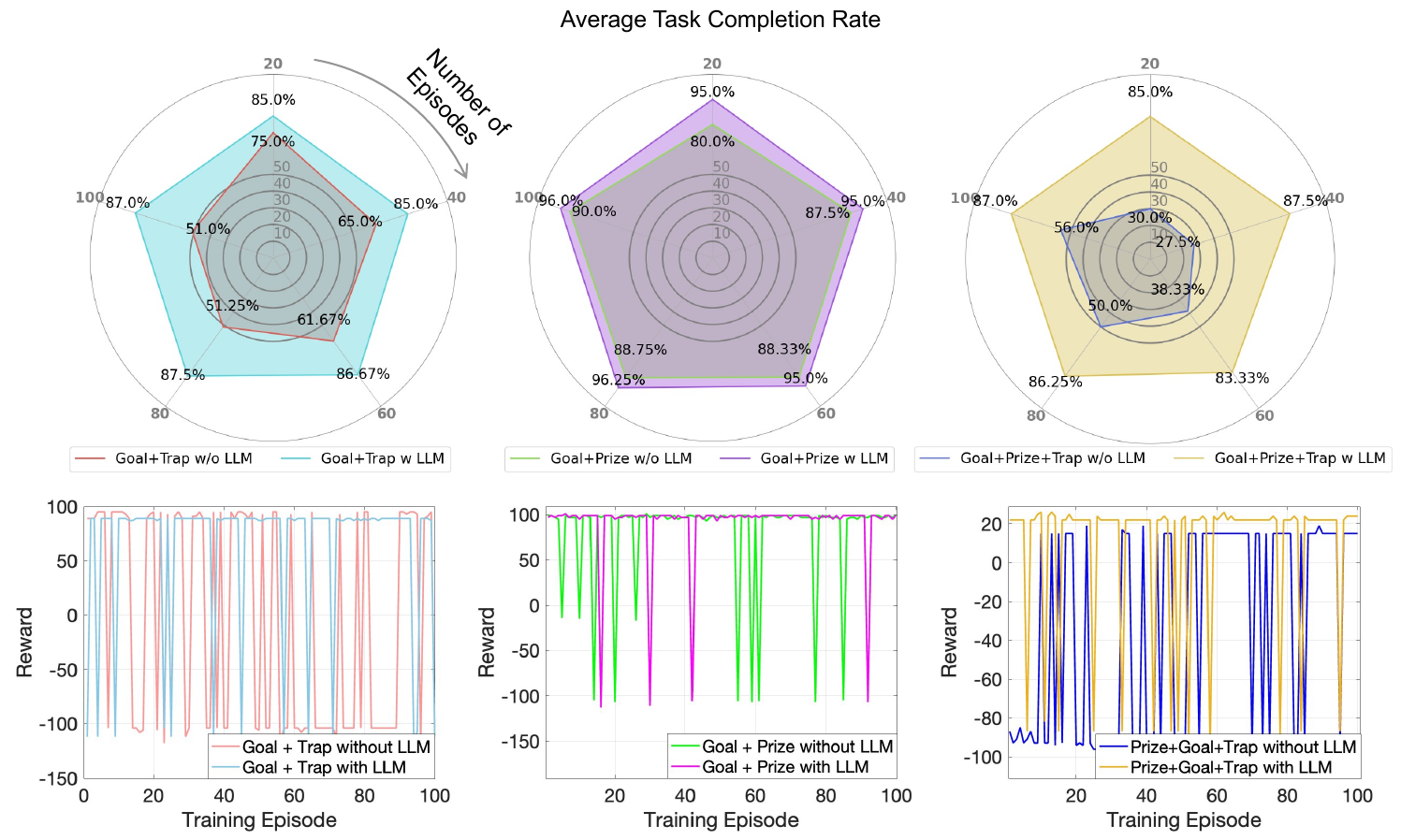}
\caption{Mixture of mission experts rewards and mission completion rate comparison.}
\label{sucessfulrate}
\end{figure*}
For the maze navigation task, we consider a walker in a grid-based simulation across three user requirements: {\textit{Goal+Trap}} (reaching the designated goal while avoiding traps), {\textit{Goal+Prize}} (acquiring a prize before reaching the goal), and {\textit{Goal+Prize+Trap}} (a combination requiring prize collection, trap avoidance, and goal attainment). 
These tasks tested the system management model's ability to integrate multiple objectives, using success rates and path efficiency as metrics. We compared a traditional gate network-enabled MoE approach and our proposed LLM-enabled MoE approach\footnote{The LLM model used in our experiments is gpt-3.5-turbo-1106 by Open AI.}.

As presented in Fig. \ref{sucessfulrate}, the gate network-enabled MoE starts with a $75\%$ success rate. However, the rate decreases because the walker tries to explore more efficient paths, increasing the risk of task failure.
Conversely, the LLM-enabled MoE maintained success rates above $85\%$, demonstrating superior strategic balance and decision-making.
In the more straightforward Goal+Prize task, the gate network MoE's performance linearly increases due to the straightforward objective. However, in the challenging Goal+Prize+Trap mission, the gate network MoE starts at a $30\%$ success rate, while the LLM-enabled MoE starts at $85.5\%$, showing LLM's adeptness in handling complex situations. Furthermore, the LLM-enabled MoE completes missions more efficiently with fewer steps and requires less intricate reward strategies, highlighting its adaptability.

\subsection{NSP Utility Maximization Task}\label{fheliah}
\begin{figure*}[t]
\centering
\includegraphics[width = 0.85\textwidth]{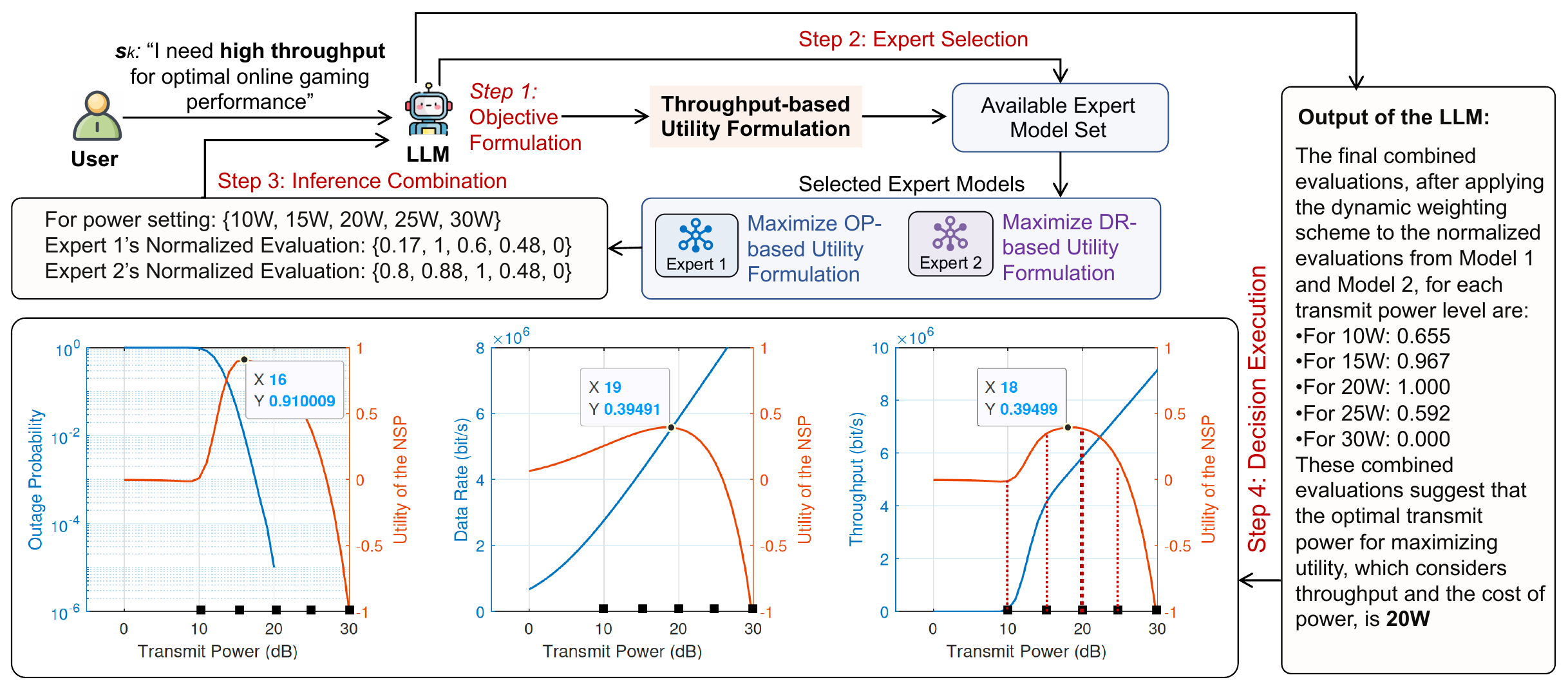}
\caption{NSP's utility under diverse user requirements and the variations of outage probability, data rate, and throughput with transmit power.}
\label{confshow}
\end{figure*}
Fig.~\ref{confshow} presents the LLM-enabled MoE workflow, illustrating that the NSP's optimal transmit power is adaptive to user-specific requirements. We consider $M =10$, $\theta = 6$, $D = 10$, $\alpha = 2$, $\sigma = 1$, $\beta_1 = 1$, $\beta_2 = 0.003$, $B = 10^6$, and $\gamma_{\rm th} = 10$ ${\rm dB}$.
The decision-making process is shown in four steps. 
Step 1 translates the user's demand for optimal online gaming into a throughput-based utility formulation. 
Step 2, {\textit{`Expert Selection,'}} involves choosing from available expert models. Step 3, {\textit{`Inference Combination,'}} shows the normalized evaluations of two selected experts for different power settings. We can observe that Expert 1 prefers a transmit power setting of $15$ W, whereas Expert 2 proposes a setting of $20$ W.
In Step 4, the LLM integrates the decisions of two expert models by examining the characteristics of OP and DR. The LLM assesses the power setting evaluations of two expert models to identify a scheme that balances throughput needs with energy cost, where the NSP's utility is maximized.

\section{Conclusion}
We introduced an innovative LLM-enabled MoE framework to address network optimization challenges in the context of diverse user requirements. LLM-enabled MoE is achieved by dynamically selecting and integrating the most appropriate expert models based on the specific demands of each user task, thereby reducing the need to train new AI models for each unique problem. The effectiveness and efficiency of LLM-enabled MoE were demonstrated through empirical testing on a maze navigation task and an NSP utility maximization task, showing its practical applicability and adaptability to complex networking systems. The results indicate a promising direction for future research in intelligent networking, where the synergy between LLMs and MoE can lead to more sustainable and user-centric network optimization solutions.

\begin{appendices}
\section{Proof of Proposition \ref{prop1}}\label{app1}
\renewcommand{\theequation}{A-\arabic{equation}}
\setcounter{equation}{0}
The OP is defined as the probability that the received SNR falls below a given threshold ${\gamma _{{\rm{th}}}}$. Thus, the OP can be expressed as ${\rm{OP}} = {F_{{\rm{SNR}}}}({\gamma _{{\rm{th}}}}){\rm{ = }}\int_0^{{\gamma _{{\rm{th}}}}} {{f_{{\rm{SNR}}}}} \left( z \right) {\rm{d}}z$. 
With the help of \eqref{PDFeq}, we obtain
\begin{equation}\label{faeoj}
{\rm{OP}} = \frac{{{{\left( {\frac{{{\sigma ^2}}}{{P{D^{ - \alpha }}}}} \right)}^M}}}{{\Gamma \left( M \right){\theta ^M}}}\int_0^{{\gamma _{{\rm{th}}}}} {{z^{M - 1}}{e^{ - \frac{{{\sigma ^2}z}}{{\theta P{D^{ - \alpha }}}}}}} {\rm{d}}z
\end{equation}
According to~\cite[eq. (8.381.8)]{gradshteyn2007}, the integral part in OP can be solved as
\begin{equation}\label{inaeikf}
\int_0^{{\gamma _{{\rm{th}}}}} {{z^{M - 1}}{e^{ - \frac{{{\sigma ^2}z}}{{\theta P{D^{ - \alpha }}}}}}} {\rm{d}}z = \frac{{\Gamma \left( {M,\frac{{{\sigma ^2}}}{{\theta P{D^{ - \alpha }}}}{\gamma _{{\rm{th}}}}} \right)}}{{{{\left( {\frac{{{\sigma ^2}}}{{\theta P{D^{ - \alpha }}}}} \right)}^M}}}.
\end{equation}
Substituting \eqref{inaeikf} into~\eqref{faeoj}, we obtain~\eqref{opequation}.

\section{Proof of Proposition \ref{prop2}}\label{app2}
\renewcommand{\theequation}{B-\arabic{equation}}
\setcounter{equation}{0}
The DR is defined as ${\rm{DR}} = B\int_0^\infty  {{{\log }_2}(1 + z){f_{{\rm{SNR}}}}} \left( z \right){\rm{d}}z$. With the help of \eqref{PDFeq}, we have
\begin{equation}\label{feijfl}
{\rm{DR}} = \frac{B}{{\Gamma \left( M \right)}}{\left( {\frac{{{\sigma ^2}}}{{\theta P{D^{ - \alpha }}}}} \right)^M}{I_{B_1}},
\end{equation}
where ${I_{B_1}} = \int_0^\infty  {{{\log }_2}\left( {1 + z} \right){z^{M - 1}}{e^{ - \frac{{{\sigma ^2}z}}{{\theta P{D^{ - \alpha }}}}}}} {\rm{d}}z$.
According to~\cite[eq. (01.04.07.0002.01)]{web}, we have
\begin{equation}\label{logerl}
{\log _2}\left( {1 + z} \right) = \frac{1}{{2\pi i}}\int_{\cal L} {\frac{{\Gamma \left( {s + 1} \right){\Gamma ^2}\left( { - s} \right){z^{ - s}}}}{{\Gamma \left( {1 - s} \right)}}} {\rm{d}}s,
\end{equation}
where $i = \sqrt { - 1}$, the integration path of $\mathcal{L}_1$ goes from $\sigma_L -i\infty $ to $\sigma_L+i\infty $ and $\sigma  \in \mathbb{R}$.
Substituting~\eqref{logerl} into $I_{B_1}$ and exchanging the
order of integration, we have
\begin{equation}
{I_{{B_1}}} = \frac{1}{{2\pi i}}\int_{\cal L} {\frac{{\Gamma \left( {s + 1} \right){\Gamma ^2}\left( { - s} \right)}}{{\Gamma \left( {1 - s} \right)}}} {I_{{B_2}}}{\rm{d}}s,
\end{equation}
where ${I_{{B_2}}} = \int_0^\infty  {{z^{M - s - 1}}{e^{ - \frac{{{\sigma ^2}z}}{{\theta P{D^{ - \alpha }}}}}}} {\rm{d}}z$.
By using~\cite[eq. (8.381.4)]{gradshteyn2007}, ${I_{{B_2}}}$ can be solved as
\begin{equation}
{I_{{B_2}}} = {\left( {\frac{{{\sigma ^2}}}{{\theta P{D^{ - \alpha }}}}} \right)^{s - M}}\Gamma \left( {M - s} \right).
\end{equation}
Substituting~${I_{{B_1}}}$ and ${I_{{B_2}}}$ into~\eqref{feijfl}, we have
\begin{equation}\label{gfaeg2}
{\rm{DR}} = \frac{{B{\Gamma ^{ - 1}}\!\left( M \right)}}{{2\pi i\ln \left( 2 \right)}}\int_{\cal L} {\frac{{\Gamma\!\left( {s + 1} \right){\Gamma ^2}\!\left( { - s} \right)}}{{\Gamma \!\left( {1 - s} \right){\Gamma ^{ - 1}}\!\left( {M - s} \right)}}} {\left( {\frac{{{D^\alpha }{\sigma ^2}}}{{\theta P}}} \right)^s}{\rm{d}}s.
\end{equation}
Using~\cite[eq. (9.301)]{gradshteyn2007}, we can rewrite~\eqref{gfaeg2} as \eqref{drequation}.

\end{appendices}

\bibliographystyle{IEEEtran}
\bibliography{Ref}

\end{document}